\documentclass[a4paper]{IEEEtran}

\usepackage{balance}

\usepackage{graphicx} 

\usepackage[english]{babel}
\usepackage[utf8]{inputenc}
\usepackage{mathtools}
\usepackage{tikz}
\usepackage{amsmath,amssymb,amsfonts}
\usepackage{pdfpages}

\usepackage{textcomp}
\usepackage{xcolor}
\def\BibTeX{{\rm B\kern-.05em{\sc i\kern-.025em b}\kern-.08em
    T\kern-.1667em\lower.7ex\hbox{E}\kern-.125emX}}

\usepackage{caption}

\usepackage{afterpage}
\usepackage{bm}
\usepackage{color}
\usepackage[font=sc]{caption}
\usepackage{floatrow}
\usepackage{listing}
\usepackage{listings}
\usepackage{fancyhdr}
\usepackage{fancyvrb} 
\usepackage{float}
\usepackage{framed}
\usepackage{enumerate}
\usepackage[shortlabels]{enumitem}
\usepackage{multirow}
\usepackage{hhline}
\usepackage{array}
\usepackage{epstopdf}
\usepackage{pgfplots}
\usepackage{subfigure}
\usetikzlibrary{matrix,chains,positioning, decorations.pathreplacing,arrows,plotmarks}
\usetikzlibrary{shapes}
\usetikzlibrary{calc} 
\pgfplotsset{compat=newest}

\usetikzlibrary{shapes.geometric, arrows}
\tikzstyle{db} = [rectangle, minimum width=0.4\columnwidth, minimum height=0.2\columnwidth, text centered, draw=black, fill=blue!20]
\tikzstyle{file} = [rectangle, rounded corners, minimum width=0.3\columnwidth, minimum height=0.1\columnwidth, text centered,
draw=black, fill=green!20]
\tikzstyle{dir} = [circle, minimum size=0.2\columnwidth, fill=red!20, draw=black]
\tikzstyle{arrow} = [thick, ->, >=stealth]
\tikzstyle{user} = [rectangle, minimum width=0.2\columnwidth, minimum height=0.05\columnwidth, draw=black, fill=black!20]

\usepackage[ruled, vlined]{algorithm2e}

\IEEEoverridecommandlockouts 
\usepackage[OT4,T1]{fontenc}
\usepackage{url}
\usepackage{epstopdf}

\usepackage{cite} 

\usepackage{balance}

\newcommand\MyBox[2]{
  \fbox{\lower0.75cm
    \vbox to 1.7cm{\vfil
      \hbox to 1.7cm{\hfil\parbox{1.4cm}{#1\\#2}\hfil}
      \vfil}%
  }%
}

\title{A Recurrent Neural Network Approach to the Answering Machine Detection Problem}
\author{
\IEEEauthorblockN{Kemal Altwlkany $^{1,2}$, Sead Delalić $^{1,2}$, Elmedin Selmanović $^{1}$, Adis Alihodžić $^{1}$, Ivica Lovrić $^{3}$
}

\IEEEauthorblockA{
$^1$Faculty of Science, University of Sarajevo, Bosnia and Herzegovina\\
$^2$Infobip, Sarajevo, Bosnia and Herzegovina\\
$^3$Infobip, Zagreb, Croatia\\
  \{kemal.altwlkany, delalic.sead, eselmanovic, adis.alihodzic\}@pmf.unsa.ba, ivica.lovric@infobip.com
}
}

\begin{document}
\maketitle          

\begin{abstract}
In the field of telecommunications and cloud communications, accurately and in real-time detecting whether a human or an answering machine has answered an outbound call is of paramount importance. This problem is of particular significance during campaigns as it enhances service quality, efficiency and cost reduction through precise caller identification. Despite the significance of the field, it remains inadequately explored in the existing literature. This paper presents an innovative approach to answering machine detection that leverages transfer learning through the YAMNet model for feature extraction. The YAMNet architecture facilitates the training of a recurrent-based classifier, enabling real-time processing of audio streams, as opposed to fixed-length recordings. The results demonstrate an accuracy of over 96\% on the test set. Furthermore, we conduct an in-depth analysis of misclassified samples and reveal that an accuracy exceeding 98\% can be achieved with the integration of a silence detection algorithm, such as the one provided by FFmpeg.
\end{abstract}

\providecommand{\keywords}[1]
{
  \textbf{\textit{Keywords---}} #1
}


\section{Introduction}
\IEEEoverridecommandlockouts\IEEEPARstart{D}{etecting} whether an outbound call was answered by a human or an answering machine (voice mail) is a common service offered by telecommunications and cloud communications platforms. This feature enables their users to be more efficient, primarily in terms of cost reduction and increasing the engagement of the user's own clients. For example, a marketer can pre-record a campaign or advertisement message and then play that message during a call with its target audience. Modern APIs by cloud communication platforms allow the marketer to initiate a bulk of these calls within clicks. However, simply playing the message to all clients is cost inefficient, as the marketer will have to pay for the entire length of the call, regardless of whether the call was answered by a human or a voice mail. Although less frequently, other undesirable situations for the marketer can occur as well, such as the call being accidentally answered while the person is not actually present in the call, e.g. due to the phone being in a pocket or placed screen facing down on a hard surface. If the marketer could know, preferably as soon as possible, whether the call was answered by a real human, the marketer could only play the message to human answered calls, hanging up in other cases. The problem being described is known as the Answering Machine Detection (AMD) problem. In its simplest form, it can be considered a binary classification problem: based on the audio content of a call (and sometimes call metadata) determine whether a human or machine has answered the call.

The work of Minaj \cite{minaj2020optimization} describes the application of AMD in call centers. Namely, it is in the interest of call centers to complete as many calls as possible during a certain period of time. The assignation of operators in call centers can be delegated to an automatic dialer. If the automatic dialer has a means of solving the AMD problem, it could skip calls answered by machines, thereby increasing the effectivness of the call center operators.

As will be shown in this paper, most cloud communications platforms offer a solution to AMD as part of their proprietary services. However, there is not much research on the given topic. The main contributions of this paper are:
\begin{itemize}
    \item Provide an overview of the current state of AMD solutions, both proprietary and open source.
    \item Offer a new approach to solving the AMD problem by applying a recurrent neural network trained using transfer learning.
    \item Enable support for modern AMD features which are currently only available from proprietary AMD providers.
    \item  Utilize a silence detection algorithm to improve the accuracy of AMD.
\end{itemize}

This paper is organized as follows: In Section 2, a review of existing solutions to AMD is provided, including both proprietary software and research advances. Section 3 provides a detailed review of the approach used in this paper. Section 4 provides insight into the results accompanied with a thorough discussion. The paper is concluded in Section 5 by providing some final remarks and suggestions for future work and improvements.

\section{Related work}

Work related to this paper can be grouped into two categories: proprietary solutions and research papers. Proprietary solutions do not reveal the underlying algorithms or techniques used, but are important as they resemble industry demands and must implement features that customers request. Research papers reveal what the current state-of-the-art is and transparently share which techniques have been applied, they provide concrete results as well as recommendations for future work. Therefore it is of importance to consider both aspects.

\subsection{Proprietary Solutions}
The introductory section defined a simple variant of the AMD problem, distinguishing between two classes: \textit{machine} and \textit{human}. Some telecommunication providers enable differentiating between more classes, thereby formulating the problem as a multiclass classification problem.

The solution provided by Infobip \cite{infobipvoiceapi} introduces two additional classes: \textit{noise} and \textit{silence}. Infobip states that their solution works best for Spanish and Portuguese languages, in countries like Spain, Colombia, Peru, Brazil and similar, achieving an accuracy of close to 95\%. For other markets, their model achieves an accuracy of roughly 80\%, but the company states that they are constantly working on improving their model. The model itself uses a detection time of 4 seconds.

The Telynx API \cite{telnyxapi} provides several AMD features, including beep detection and a premium AMD feature which performs speech recognition as well. Telnyx differentiates between three classes: \textit{human}, \textit{machine} and \textit{not sure}.

Sinch \cite{sinchapi} offers two forms of AMD: \textit{synchronous} and \textit{asynchronous}. The synchronous form simply informs the user that the detection has ended using Sinch's eventing system, while the asynchronous form also detects whether a beep has occured - enabling the user to leave a message for the callee.

Twilio \cite{twilioapi} also provides an AMD solution, taking on average around 4 seconds to complete the detection and reaching above 90\% accuracy in the US. Twilio's solution is distinct as it allows tuning parameters which allow increasing or decreasing the time needed to perform the detection, as well as other parameters, ultimately allowing the user to control their desired trade-off between speed and accuracy.

Convoso \cite{convoso} claim that their solution achieves an accuracy of 97\%, while allowing the user to tune the model to create more false positives or false negatives, i.e. depending on the use case the user might prefer a different trade-off between specificity and sensitivity of the model.

LumenVox \cite{lumenvox} provides an AMD service as well, but bundles it within their Call Progress Analysis tool, enabling detection of whether, and when, a human or machine has answered a call.

\subsection{Research Papers}
Unlike the vast number of proprietary solutions available, publicly available research regarding AMD is quite limited.
The aforementioned work of Minaj \cite{minaj2020optimization} is a master's thesis that focused on optimizing the existing AMD of Asterisk by fine-tuning the default parameters that Asterisk provides. Asterisk \cite{van2007asterisk,madsen2011asterisk} is an open-source PBX and telephony toolkit that offers many telecommunication services: conference bridges, voicemail, interactive voice responses, configuring dial plans, fax support, etc. It provides an AMD application as well, which is based on timing patterns. The AMD solution can distinguish between three classes: \textit{human}, \textit{machine} and \textit{not sure}, but can also provide a fourth class, \textit{hangup}, in case that the call is hung up by the callee. A detailed explanation of how the Asterisk algorithm works can be obtained from the official documentation. In short, the algorithm contains a specific set of rules which classify the call. For example, a parameter called \textit{greeting} refers to the maximum length that an initial greeting can last. Should the greeting time exceed the value of the parameter, the audio is classified as a \textit{machine} \cite{van2007asterisk}. Examples of other parameters are \textit{minimum word length} - referring to the minimum length of vocal activity for it to be considered a word or \textit{maximum number of words} - referring to the maximum number of words which can be detected, which if exceeded result in classifying the audio as \textit{machine}. This is an explicitly programmed solution which does not employ any machine learning techniques. The main drawback is that the only differentatior between silence and words in Asterisk is given by the silence/noise threshold \cite{van2007asterisk,gomez2014answering}, as there is no advanced voice activity detection mechanism.

Gomez et al. \cite{gomez2014answering} provide an AMD system which labels audio as human or machine within three seconds. To form their prediction, they use three main sources of information: time required to pick up the phone, the percentage of silence within the first three seconds and the recognized word sequence. The entire module which forms the prediction is referred to as the \textit{detector} and is integrated into Asterisk. The detector separately computes three probabilities, based on the previously mentioned three sources of information, each of which indicate the probability that the call was answered by a human. Finally, based on their available data, they derived a formula for computing the joint probability that the overall call was answered by a human. It is worth noting that the most complex part of their algorithm, the Automatic Speech Recognition algorithm (ASR) is a Large Vocabulary Continuous Speech Recognizer (LVCSR) with a predefined and reduced lexicon just containing the words of expected sentences.

To this day, the authors of this paper have found that the work of Anisomov et al. \cite{anisimovanswering} is the only publicly available research paper that solves the AMD using a deep learning approach. In their paper, a convolutional neural network (CNN) is trained to differentiate between two classes: \textit{human} and \textit{machine}. As input to their model, they trim raw audio to exactly 2.5 seconds of length and compute the Mel frequency cepstral coefficients (MFCC). Extracting MFCCs and using them as compressed representations of audio is a fairly common technique in audio processing \cite{fayek2016speech,abdul2022mel,ayvaz2022automatic,muda2010voice}. After computing the MFCCs, the authors discard the first cepstral coefficient, as it is a measure of signal loudness/DC component of the signal \cite{anisimovanswering,fayek2016speech}. The MFCC were computed per every 62.5 milliseconds of audio, resulting in a total of 40 frames. The final input to their CNN is a 59 by 40 matrix, whereas a sigmoid function was used in the output layer. The authors claim to achieve an accuracy of 94\%, which corresponds to industry standard \cite{anisimovanswering}.
\section{Case Study}
In Section 2, it was outlined that publicly available research papers on AMD are scarce. Existing proprietary solutions offer more features and flexibility than publicly available solutions. The goal is to close this gap and deliver a new publicly available AMD method.

Any new solution should be easy to extend so that it can support additional features, such as detecting the end of a message, or enabling the user to tune the trade-off between specificity and sensitivity. In modern times, deep learning is a promising research area which can scale incredibly well with large data, thus the solution will be implemented using deep learning.

AMD is used for live calls, therefore the entire audio is not available immediately, instead it is constantly being streamed over a media exchange protocol such as RTP \cite{rfc3550}. Existing AMD solutions do not leverage the streaming aspect of live calls, instead they wait until the required amount of call audio has been accumulated and then perform inference. Utilizing the real-time streaming aspect of AMD is desirable, as it could lead to faster performances, with better time localization and will make extending the solution to support additional features easier.

What follows is an in-depth overview of the newly proposed AMD solution. With transfer learning, YAMNet \cite{tensorflowmodelgarden2020} is used to extract embedddings from audio and adapted to work in a stream-like fashion. A recurrent classifier is then trained to classify the stream as \textit{human} or \textit{machine}. Suggestions on incorporating additional parameters which enable behaviour similar to that of proprietary solutions is provided. Special attention is paid to how inference and model serving should be done regarding streaming.

\subsection{Model Architecture}

\subsubsection{Feature extraction}
YAMNet \cite{tensorflowmodelgarden2020} was used to facilitate a transfer learning approach rather than manually extracting the features. YAMNet is a deep learning model trained on the AudioSet \cite{audioset} dataset that can classify audio into 521 classes. Since AMD must be solved in real-time, YAMNet is a good option given that it employs the Mobilenet\_v1 depth-wise separable convolution architecture \cite{tensorflowmodelgarden2020,howard2017mobilenets}. MobileNets are an efficient class of models which provide a good trade off between latency and accuracy. However, YAMNet itself is not designed to be used directly in a stream-like fashion. If given an audio of arbitrary duration, YAMNet extracts \textit{frames} from it that are exactly 960 milliseconds long and it extracts one frame every 480 milliseconds. YAMNet will classify each frame individually, i.e. each frame is assigned a label.

The 480 millisecond stride performed by YAMNet can be adapted to support streaming. Namely, after an initial 480 milliseconds of audio stream have been accumulated, the model can be fed new data every 480 milliseconds concatenated with the last 480 milliseconds replicating a real-time stride, or the parameter known as \textit{hop length} used when generating spectrograms \cite{mcfee2015librosa}. A buffer of 960 milliseconds of audio seems appropriate given that most voice codecs operate on packets varying from 5 to 60 milliseconds, most commonly 10 and 20 milliseconds \cite{ITUrecG711,ITUrecG729,rfc6716,vos-silk-02}.

The convolutional approach to AMD proposed by Anisimov et al. relied on MFCCs \cite{anisimovanswering}. Internally, YAMNet creates a stabilized logmel spectrogram \cite{tensorflowmodelgarden2020} before extracting the features, but both features are common in state-of-the-art deep learning for speech processing, e.g. Whisper, OpenAI's speech-to-text model, and Pretrained Audio Neural Networks, both use the log mel spectrogram \cite{radford2023robust,kong2020panns}.

\subsubsection{GRU classifier}
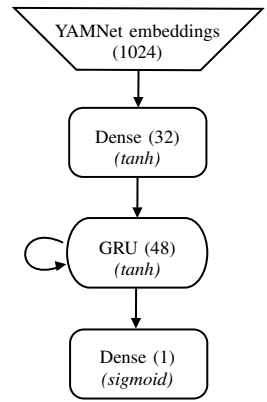
\begin{figure}[H]
\captionsetup{font=small}
    \centering
    \scalebox{1.0}{

\tikzset{every picture/.style={line width=0.75pt}} 

\begin{tikzpicture}[x=0.75pt,y=0.75pt,yscale=-1,xscale=1]

\draw   (102.11,124.8) -- (149.71,124.8) .. controls (155.9,124.8) and (160.91,132.64) .. (160.91,142.3) .. controls (160.91,151.96) and (155.9,159.8) .. (149.71,159.8) -- (102.11,159.8) .. controls (95.93,159.8) and (90.91,151.96) .. (90.91,142.3) .. controls (90.91,132.64) and (95.93,124.8) .. (102.11,124.8) -- cycle ;
\draw    (88.92,137.22) .. controls (65.26,125.58) and (62.45,158.78) .. (88.07,149.05) ;
\draw [shift={(90.52,148.02)}, rotate = 169.51] [fill={rgb, 255:red, 0; green, 0; blue, 0 }  ][line width=0.08]  [draw opacity=0] (5.36,-2.57) -- (0,0) -- (5.36,2.57) -- cycle    ;
\draw   (91.96,76.33) .. controls (91.96,72.49) and (95.07,69.38) .. (98.91,69.38) -- (153.51,69.38) .. controls (157.35,69.38) and (160.46,72.49) .. (160.46,76.33) -- (160.46,97.18) .. controls (160.46,101.02) and (157.35,104.13) .. (153.51,104.13) -- (98.91,104.13) .. controls (95.07,104.13) and (91.96,101.02) .. (91.96,97.18) -- cycle ;
\draw   (92.4,187.8) .. controls (92.4,183.96) and (95.51,180.85) .. (99.35,180.85) -- (153.95,180.85) .. controls (157.79,180.85) and (160.9,183.96) .. (160.9,187.8) -- (160.9,208.65) .. controls (160.9,212.49) and (157.79,215.6) .. (153.95,215.6) -- (99.35,215.6) .. controls (95.51,215.6) and (92.4,212.49) .. (92.4,208.65) -- cycle ;

\draw    (126.07,104.64) -- (125.99,121.79) ;
\draw [shift={(125.98,124.79)}, rotate = 270.24] [fill={rgb, 255:red, 0; green, 0; blue, 0 }  ][line width=0.08]  [draw opacity=0] (5.36,-2.57) -- (0,0) -- (5.36,2.57) -- cycle    ;
\draw    (125.36,160.29) -- (125.28,177.43) ;
\draw [shift={(125.27,180.43)}, rotate = 270.24] [fill={rgb, 255:red, 0; green, 0; blue, 0 }  ][line width=0.08]  [draw opacity=0] (5.36,-2.57) -- (0,0) -- (5.36,2.57) -- cycle    ;
\draw   (95.57,50.21) -- (64.45,19.1) -- (188.43,19.57) -- (157.55,50.45) -- cycle ;
\draw    (126.57,50.71) -- (126.45,66.29) ;
\draw [shift={(126.43,69.29)}, rotate = 270.44] [fill={rgb, 255:red, 0; green, 0; blue, 0 }  ][line width=0.08]  [draw opacity=0] (5.36,-2.57) -- (0,0) -- (5.36,2.57) -- cycle    ;

\draw (126,201) node  [font=\scriptsize] [align=left] {\begin{minipage}[lt]{68pt}\setlength\topsep{0pt}
\begin{center}
Dense (1)\\\textit{(sigmoid)}
\end{center}

\end{minipage}};
\draw (126,146) node  [font=\scriptsize] [align=left] {\begin{minipage}[lt]{68pt}\setlength\topsep{0pt}
\begin{center}
GRU (48)\\\textit{(tanh)}
\end{center}

\end{minipage}};
\draw (126,90) node  [font=\scriptsize] [align=left] {\begin{minipage}[lt]{68pt}\setlength\topsep{0pt}
\begin{center}
Dense (32)\\\textit{(tanh)}
\end{center}

\end{minipage}};
\draw (126,36.5) node  [font=\scriptsize] [align=left] {\begin{minipage}[lt]{68.39pt}\setlength\topsep{0pt}
\begin{center}
YAMNet embeddings (1024)
\end{center}

\end{minipage}};

\end{tikzpicture}

}
    \caption{Model architecture}
    \label{fig::model architecture}
\end{figure}

A combination of GRU layers and Dense layers was used to build a classifier on top of YAMNet which will classify audio into two classes: \textit{human} and \textit{machine}. The inputs to the classifier will be 1024-dimensional embeddings that YAMNet generates, per every frame (960 milliseconds with a 480 millisecond stride). Since YAMNet is a powerful network that can distinguish between 521 different classes with an excellent mean average precision (mAP) score of 0.306 \cite{tensorflowmodelgarden2020}, the GRU classifier does not require a large depth nor many parameters - it is expected that the embeddings extracted by YAMNet are of high quality. Since this is a binary classifiction problem, a simple sigmoid was chosen as the activation function of the last layer. After some experimentations with the hyperparameters, a smaller grid search was performed and the architecture provided by Fig. \ref{fig::model architecture} was selected.

As can be seen immediately from Fig. \ref{fig::model architecture}, the network is relatively shallow, excluding the feature extraction layers of YAMNet. The model has only 44.657 trainable parameters in the classifier part, but due to the quality of the features extracted by YAMNet, it can still achieve state-of-the-art accuracy.

\subsection{Training}
The model was implemented using the Keras framework \cite{chollet2015keras}. The training dataset contains around 4200 audio files and was provided by Infobip. The dataset is well balanced between the two classes of interest and a train-test-validation split was created, as shown in Table \ref{tab::data split}. To speed up training and enable a batched training process, all audio files were clipped or zero-padded to be of equal duration. The reason for this is the limitation of Keras, because the training process cannot be parallelized if the number of timesteps is not equal in a single batch \cite{chollet2015keras}. The chosen duration was 4 seconds since most files in the dataset were a little over that duration, thereby changes were avoided as much as possible.

Regarding other relevant hyperparameters, mostly default values were used: binary crossentropy as the loss function, Adam \cite{kingma2014adam} as the optimizer and a per-epoch callback was implemented that memorized model weights with best accuracy on the validation set, instead of the weights of the ultimate epoch.

\begin{table}[H]
\center
\begin{tabular}{|l|l|l|l|}
\cline{2-4}
\multicolumn{1}{l|}{}
        & Train & Test & Validation \\ \hline
Human   & 1679  & 210  & 210        \\ \hline
Machine & 1732  & 215  & 216        \\ \hline
\end{tabular}
\caption{Dataset split and class distribution}
\label{tab::data split}
\end{table}

\subsection{Inference}
\subsubsection{Streaming the input}
Instead of accumulating a predefined duration of audio (e.g. 2.5 seconds) and running inference once, the AMD architecture proposed in this paper enables running inference virtually an arbitrary amount of times. The process is similar to that of generating a live, real-time spectrogram. After an outbound call has connected, audio will start streaming from the callee's side. This audio is accumulated into a buffer on the caller's side. After the buffer has accumulated enough audio content for a frame, the frame from the buffer is passed to the AMD model in order to run inference. Naturally, the frame duration was selected to match that of YAMNet, i.e. 960 milliseconds.

Even after such a small frame of audio, the end-user can be provided with a result. As can be expected, such a result will likely not be that accurate, as 960 milliseconds is rarely enough to determine whether a call was answered by a human, or a machine. Thus, the process can continue, in real-time.

After another 480 milliseconds of audio have been streamed and put into the buffer, the last 960 milliseconds of audio are retrieved from the buffer and passed to the model for inference. A new inference result is obtained, and the process described above is repeated in an iterative fashion.

\subsubsection{Additional parameters and termination conditions}
The process described above could run indefinitely, unless the streaming is stopped (e.g. the call is hung up) or unless a termination condition is introduced. Instead of fixating a termination condition, parameters which define the termination conditions can be exposed to the end-user. For example, a \textit{timeout} value can be defined that would terminate the AMD process and notify the end-user if AMD exceeds the duration defined by \textit{timeout}.

Another useful parameter is \textit{confidence threshold}. Since the output of the classifier is a sigmoid, its value lies between 0 and 1, for human and machine respectively. Thus, the closer the output of the network is to 0, the more confident the AMD is that the call was answered by a human. The same applies for value 1 and confidence regarding a machine. This enables the end-user to define a \textit{confidence threshold}, which once met would stop the AMD process.

The downside of introducing a \textit{confidence threshold} is that sometimes the model can be confident in the very first few frames, even though it might change its opinion afterwards. As a counter-measure parameter
\textit{minimum detection time} is introduced. This parameter does not enable the AMD to stop earlier, regardless of the confidence threshold, unless at least \textit{minimum detection time} has elapsed. Together with the previous parameters, these three values enable the end-user to fine-tune the behaviour of AMD and decide between an on average faster but less confident, or slower
but more reliable AMD.

\subsection{Model Serving and Statefulness}
The benefits of streaming come with a price to pay - a more complex architecture for serving the model is required. Once a call has been established the entire audio will not be available momentarily, but is streamed into the model frame by frame. Two propositions on solving this problem are given:

\begin{enumerate}
    \item Use a stateful model. In this scenario, the classifier model in Keras must have the parameter \textit{stateful} set to true \cite{chollet2015keras}. This ensures that once inference has been completed for one frame, Keras will not reset the internal states of the recurrent layers, but use them as initial values for the next inference/frame. However, this prevents using the same model for parallel calls, as the states should not be shared among them. Instead, a single model must be attached to a single call. This solution is simpler to implement and will perform faster, as no redundant computations are performed. However, the scalability of such a solution is incredibly poor: an entire model must be allocated per every ongoing parallel call.
    \item Memorize the embeddings. In this case, YAMNet and the classifier are kept as separate entities and can be shared among parallel calls. Upon establishing a call, the first frame is passed to YAMNet. YAMNet will extract the embeddings which are then passed to the GRU classifier. The embeddings are also cached in a separate entity. Upon accumulating a new frame, only the new frame is passed to YAMNet. The newly extracted features are then appended to the already extracted features of the first frame into the cache. Together, they are passed to the GRU classifier. The model will perform a redundant computation in this case, as it has already classified the first frame in the previous inference. Since statefulness of models is not required in this case, a high-performance machine learning serving system TensorFlow-Serving \cite{olston2017tensorflow}, specifically designed for production environments, can be used. TensorFlow-Serving can scale incredibly well, and a single model will provide faster inference times than regular Python API. It is not surprising given that Google uses TensorFlow-Serving for its machine learning services.
\end{enumerate}

The second approach for model serving is recommended. It uses a state-of-the-art serving solution which is maintained by Google. Even though redundant computations do occur, this is not that uncommon for machine learning models which require knowledge of a previous state. The popular chatbot ChatGPT \cite{brown2020language,ouyang2022training} operates in a similar manner, with Azure's and OpenAI's API's requiring the entire converation to be sent upon inference \cite{openaiapi,azureapi}.
\section{Results}

It should be noted that in the results section, none of the parameters which would enable additional features (\textit{timeout}, \textit{confidence threshold} and \textit{minimum detection time}) were required, as their values do not influence the accuracy and performance of the deep learning model.

All experiments were performed on a Linux machine (Ubuntu 22.04 LTS), with a Intel Core i7-1280P processor and 32GB RAM, without the usage of a GPU. 

\subsection{Performance on Train, Test and Validation Set}
Table \ref{tab::accuracy} provides the accuracy per training, test and validation sets. The model achieves 96.67\% accuracy on the test set, which is quite good and compares to existing state-of-the-art models, both proprietary and open-source solutions.

\begin{table}[H]
\captionsetup{font=sc}
\center
\begin{tabular}{|c|c|c|c|}
\cline{2-4}
\multicolumn{1}{c|}{}
         & Train & Test  & Validation \\ \hline
Accuracy & 97.72\% & 96.67\% & 95.84\%      \\ \hline
\end{tabular}
\caption{Accuracy per dataset split}
\label{tab::accuracy}
\end{table}

Table \ref{tab::confusion matrix test set} shows the confusion matrix obtained on the test set. An interesting detail can be inferred from the confusion matrix: by summing all the cells, it appears as if the total number of samples is 3400. Obviously, as stated in table \ref{tab::data split} this is not the case. The reason is that Keras was instructed to treat every frame as an independent instance. This can quickly be verified: the total number of instances in the test set is 425. The length of each file in the test set is the same and is equal to 4 seconds. This means that each file has a total of 8 frames, as the stride performed per frame is 480 milliseconds.

The confusion matrix provides insight into per-frame inference of the AMD classifier. The actual, final output of the AMD model would require defining the additional parameters: \textit{timeout}, \textit{confidence threshold} and \textit{minimum detection time} and it would depend on the values of those parameters specified by the end-user.

\begin{table}[H]
\center

\begin{tabular}{l|l|c|c|c}
\multicolumn{2}{c}{}&\multicolumn{2}{c}{Actual}&\\
\cline{3-4}
\multicolumn{2}{c|}{}&Human&Machine&\multicolumn{1}{c}{Total}\\
\cline{2-4}
\multirow{2}{*}{Predicted}& Human & $1635$ & $45$ & $1680$\\
\cline{2-4}
& Machine & $67$ & $1653$ & $1720$\\
\cline{2-4}
\multicolumn{1}{c}{} & \multicolumn{1}{c}{Total} & \multicolumn{1}{c}{$1702$} & \multicolumn{    1}{c}{$1698$} & \multicolumn{1}{c}{$3400$}\\
\end{tabular}
\caption{Confusion matrix of test set}
\label{tab::confusion matrix test set}
\end{table}

\subsection{Improving Performance Using Silence Detection}
Manual investigation of the misclassified frames showed that a significant percentage of them contains low vocal activity or are almost completely silent. It is not uncommon for deep learning models to utilize information about vocal energy, e.g. RNNoise \cite{valin2018hybrid}, a deep learning model for speech enhancement uses a voice activity detection (VAD) module. In this paper a simpler form of VAD was used, which does not necessarily detect speech, but generally computes the decibels relative to full scale, a behaviour similar to that of FFmpeg's silence detection \cite{silencedetect}. By incorporating silence detection into AMD, frames which do not contain enough vocal energy can be skipped for inference. Table \ref{tab::silence detection acc} provides a comparison of commonly used machine learning metrics with and without the usage of silence detection. A marginal increase in sensitivity is observed, whereas a significant improvement of specificity occurred. This is due to the fact that most of the misclassified instances which were also silent frames were false negatives.

\begin{table}[H]
\center

\resizebox{\linewidth}{!}{

    \begin{tabular}{llllll}
    \multicolumn{2}{l}{\multirow{2}{*}{}}                                                                                             & \multicolumn{4}{c}{Metric}                                                                                                           \\ \cline{3-6} 
    \multicolumn{2}{l|}{}                                                                                                              & \multicolumn{1}{l|}{Accuracy} & \multicolumn{1}{l|}{Precision} & \multicolumn{1}{l|}{Specificity} & \multicolumn{1}{l|}{Sensitivity} \\ \cline{2-6} 
    \multicolumn{1}{c|}{\multirow{2}{*}{\begin{tabular}[c]{@{}c@{}}Silence\\ detection\end{tabular}}} & \multicolumn{1}{l|}{Disabled}  & \multicolumn{1}{l|}{96.71\%}    & \multicolumn{1}{l|}{96.15\%}     & \multicolumn{1}{l|}{96.01\%}       & \multicolumn{1}{l|}{97.38\%}           \\ \cline{2-6} 
    \multicolumn{1}{c|}{}                                                                             & \multicolumn{1}{l|}{Enabled} & \multicolumn{1}{l|}{98.10\%}        & \multicolumn{1}{l|}{98.54\%}         & \multicolumn{1}{l|}{98.55\%}           & \multicolumn{1}{l|}{97.65\%}           \\ \cline{2-6} 
    \end{tabular}
}
\caption{Comparison of metrics with and without silence detection}
\label{tab::silence detection acc}
\end{table}

\subsection{Inference Time}
\begin{figure}
    \centering
    \includegraphics[width=1.0\linewidth]{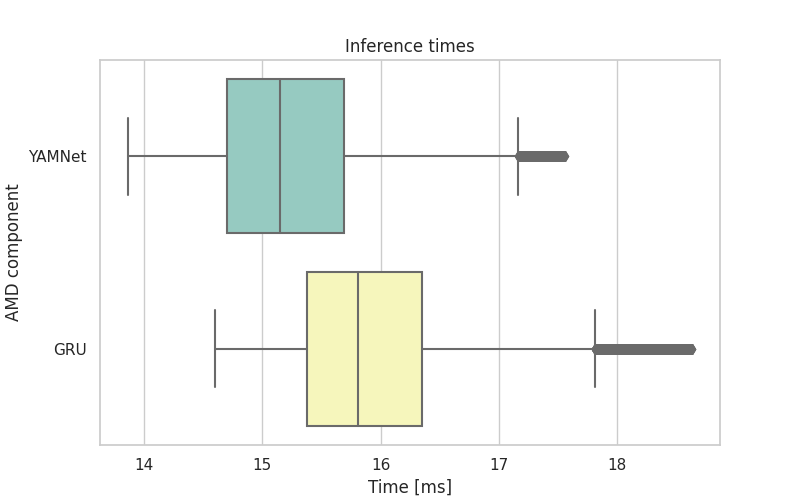}
    \caption{Inference time per YAMNet and GRU components}
    \label{fig::inference}
\end{figure}

Figure \ref{fig::inference} provides a boxplot of inference times per single frame per AMD component: the YAMNet feature extraction and the GRU classifier. This figure is an excellent example of the efficiency of the YAMNet architecture, even though the model has 3.7 million parameters \cite{tensorflowmodelgarden2020} it  performs slightly faster on average compared to the GRU classifier. The average inference time of YAMNet is 15.24 milliseconds, whereas it takes 15.93 milliseconds for the GRU component. Both values were computed while omitting the outliers: the top 2\% of quickest and slowest inference times.

The silence detection module takes on average 0.46 milliseconds per frame, which is negligible compared to the YAMNet and GRU components. Therefore it was omitted from the boxplot as it could not render clearly. Summing the three components together yields an average inference time of 31.63 milliseconds per frame, which is quite good. This is roughly a delay time of 1 to 6 audio packets for many audio codecs, a delay that can be tolerated when operating in real-time \cite{ITUrecG711,ITUrecG729,rfc6716,vos-silk-02}.

\section{Conclusion}
This paper provided a new approach to solving the AMD problem. Existing proprietary solutions from telecommunications and cloud communications platforms were analyzed and a review of existing research literature on the given topic was provided.

Given that AMD is a problem which must be solved real-time, the newly proposed solution enables utilization of streaming capabilities. Using transfer learning, a GRU classifier with a small number of parameters (44.657) was trained to distinguish between human and machine answered calls. The model provides an accuracy of 96.67\% on the test set, which was further improved to 98.10\% by using a silence detection module. Both results are comparable to state-of-the-art solutions.

The proposed solution also enables end-users to customize AMD's behaviour via three additional parameters: \textit{timeout}, \textit{confidence threshold} and \textit{minimum detection time}.

For future work, it would be interesting to provide a language/regional based analysis of the dataset.

The silence detection module could be integrated into the classifier, by adding the result of the module as an additional input to the classifier.

The currently provided results did not use any data augmentation techniques, which might be interesting to review.

Another direction for future research would be to provide a direct comparison of existing solutions on a single, representative dataset, taking into consideration both model accuracy but also inference speed and CPU/memory consumption.

The streaming capability in conjunction with the additional parameters could be used to enable end-users to leave a message after, and if, a machine was detected.

Finally, it would be beneficial to provide a case study of production-stable serving solutions for stateful models. Such a study would not be limited to AMD, but could be generalized to other problems which require models with memory and/or internal states.

\balance

\bibliographystyle{ieeetr}
\bibliography{references.bib}

\end{document}